\documentclass{amsart}
\usepackage{amssymb}
\usepackage{amsmath}
\usepackage{amsfonts}

\newtheorem{theorem}{Theorem}
\theoremstyle{plain}
\newtheorem{acknowledgement}[theorem]{Acknowledgement}

\newtheorem{definition}[theorem]{Definition}

\newtheorem{problem}[theorem]{Problem}

\newtheorem{remark}[theorem]{Remark}

\numberwithin{equation}{section}
\numberwithin{theorem}{section}
\input{tcilatex}

\begin{document}
\title[Statistical restricted isometry property]{The statistical restricted
isometry property and the Wigner semicircle distribution of incoherent
dictionaries}
\author{Shamgar Gurevich}
\address{Department of Mathematics, University of California, Berkeley, CA
94720, USA. }
\email{shamgar@math.berkeley.edu}
\author{Ronny Hadani}
\address{Department of Mathematics, University of Chicago, IL 60637, USA.}
\email{hadani@math.uchicago.edu }
\date{Dec. 1, 2008.}
\thanks{\copyright \ Copyright by S. Gurevich and R. Hadani, Dec. 1, 2008.
All rights reserved.}

\begin{abstract}
In this article we present a statistical version of the Cand\`{e}s-Tao
restricted isometry property (SRIP for short) which holds in general for any
incoherent dictionary which is a disjoint union of orthonormal bases. In
addition, we show that, under appropriate normalization, the eigenvalues of
the associated Gram matrix fluctuate around $\lambda =1$ according to the
Wigner semicircle distribution. The result is then applied to various
dictionaries that arise naturally in the setting of finite harmonic
analysis, giving, in particular, a better understanding on a remark of
Applebaum-Howard-Searle-Calderbank concerning RIP for the Heisenberg
dictionary of chirp like functions.
\end{abstract}

\maketitle

\section{Introduction}

Digital signals, or simply signals, can be thought of as complex valued
functions on the finite field $\mathbb{F}_{p},$ where $p$ is a prime number.
The space of signals $\mathcal{H=%
%TCIMACRO{\U{2102} }%
%BeginExpansion
\mathbb{C}
%EndExpansion
}\left( \mathbb{F}_{p}\right) $ is a Hilbert space of dimension $p$, with
the inner product given by the standard formula 
\begin{equation*}
\left \langle f,g\right \rangle =\tsum \limits_{t\in \mathbb{F}_{p}}f\left(
t\right) \overline{g\left( t\right) }.
\end{equation*}

A dictionary $\mathfrak{D}$ is simply a set of vectors (also called \textit{%
atoms}) in $\mathcal{H}$. The number of vectors in $\mathfrak{D}$ can exceed
the dimension of the Hilbert space $\mathcal{H}$, in fact, the most
interesting situation is when $\left \vert \mathfrak{D}\right \vert \gg
p=\dim \mathcal{H}$. In this set-up we define a \textit{resolution} of the
Hilbert space $\mathcal{H}$ via $\mathfrak{D}$, which is the morphism of
vector spaces 
\begin{equation*}
\Theta :%
%TCIMACRO{\U{2102} }%
%BeginExpansion
\mathbb{C}
%EndExpansion
\left( \mathfrak{D}\right) \rightarrow \mathcal{H}\text{,}
\end{equation*}%
given by $\Theta \left( f\right) =\sum_{\varphi \in \mathfrak{D}}f\left(
\varphi \right) \varphi $, for every $f\in 
%TCIMACRO{\U{2102} }%
%BeginExpansion
\mathbb{C}
%EndExpansion
\left( \mathfrak{D}\right) $. A more concrete way to think of the morphism $%
\Theta $ is as a $p\times \left \vert \mathfrak{D}\right \vert $ matrix with
the columns being the atoms in $\mathfrak{D}$. \ 

In the last two decades \cite{DGM}, and in particular in recent years \cite%
{BDE, C, Ca, CRT, CT, D}, resolutions of Hilbert spaces became an important
tool in signal processing, in particular in the emerging theories of
sparsity and compressive sensing.

\section{The restricted isometry property}

A useful property of a resolution is the restricted isometry property (RIP
for short) defined by Cand\`{e}s-Tao in \cite{CT}. Fix a natural number $%
n\in 
%TCIMACRO{\U{2115} }%
%BeginExpansion
\mathbb{N}
%EndExpansion
$ and a pair of positive real numbers $\delta _{1},\delta _{2}\in 
%TCIMACRO{\U{211d} }%
%BeginExpansion
\mathbb{R}
%EndExpansion
_{>0}$.

\begin{definition}
A dictionary $\mathfrak{D}$ satisfies the \underline{\textit{restricted
isometry property}} with coefficients $\left( \delta _{1},\delta
_{2},n\right) $ if for every subset $S\subset \mathfrak{D}$ such that $%
\left
\vert S\right \vert \leq n$ we have%
\begin{equation*}
\left( 1-\delta _{2}\right) \left \Vert f\right \Vert \leq \left \Vert
\Theta \left( f\right) \right \Vert \leq \left( 1+\delta _{1}\right) \left
\Vert f\right \Vert ,
\end{equation*}%
for every function $f\in 
%TCIMACRO{\U{2102} }%
%BeginExpansion
\mathbb{C}
%EndExpansion
\left( \mathfrak{D}\right) $ which is supported on the set $S$.
\end{definition}

Equivalently, RIP can be formulated in terms of the spectral radius of the
corresponding Gram operator. Let $\mathbf{G}\left( S\right) $ denote the
composition $\Theta _{S}^{\ast }\circ \Theta _{S}$ with $\Theta _{S}$
denoting the restriction of $\Theta $ to the subspace $%
%TCIMACRO{\U{2102} }%
%BeginExpansion
\mathbb{C}
%EndExpansion
_{S}\left( \mathfrak{D}\right) \subset 
%TCIMACRO{\U{2102} }%
%BeginExpansion
\mathbb{C}
%EndExpansion
\left( \mathfrak{D}\right) $ of functions supported on the set $S$. The
dictionary $\mathfrak{D}$ satisfies $\left( \delta _{1},\delta _{2},n\right) 
$-RIP if for every subset $S\subset {\footnotesize D}$ such that $%
\left
\vert S\right \vert \leq n$ we have 
\begin{equation*}
\delta _{2}\leq \left \Vert \mathbf{G}\left( S\right) -Id_{S}\right \Vert
\leq \delta _{1},
\end{equation*}%
where $Id_{S}$ is the identity operator on $%
%TCIMACRO{\U{2102} }%
%BeginExpansion
\mathbb{C}
%EndExpansion
_{S}\left( \mathfrak{D}\right) $.

It is known \cite{BDDW, D} that the RIP holds for random dictionaries.
However, one would like to address the following problem \cite{AHSC, De, DE,
HCS, I, J, JXHC, S, XH, Tr1, Tr2}:

\begin{problem}
\label{deterministic_prb}Find \underline{deterministic} construction of a
dictionary $\mathfrak{D}$ with $\left \vert \mathfrak{D}\right \vert \gg p$
which satisfies RIP with coefficients in the critical regime 
\begin{equation}
\delta _{1},\delta _{2}\ll 1\text{ and }n=\alpha \cdot p,
\label{critical_eq}
\end{equation}%
for some constant $0<\alpha <1$.
\end{problem}

\section{Incoherent dictionaries}

Fix a positive real number $\mu \in 
%TCIMACRO{\U{211d} }%
%BeginExpansion
\mathbb{R}
%EndExpansion
_{>0}$. The following notion was introduced in \cite{DE, EB} and was used to
study similar problems in \cite{Tr1, Tr2}:

\begin{definition}
A dictionary $\mathfrak{D}$ is called \underline{incoherent} with coherence
coefficient $\mu $ (also called $\mu $-coherent) if for every pair of
distinct atoms $\varphi ,\phi \in \mathfrak{D}$ 
\begin{equation*}
\left \vert \left \langle \varphi ,\phi \right \rangle \right \vert \leq 
\frac{\mu }{\sqrt{p}}\text{.}
\end{equation*}
\end{definition}

In this article we will explore a general relation between RIP and
incoherence. Our motivation comes from three examples of incoherent
dictionaries which arise naturally in the setting of finite harmonic
analysis:

\begin{itemize}
\item The first example \cite{H, HCM}, referred to as the \textit{Heisenberg
dictionary} $\mathfrak{D}_{H}$, is constructed using the Heisenberg
representation of the finite Heisenberg group $H\left( \mathbb{F}_{p}\right) 
$. The Heisenberg dictionary is of size approximately $p^{2}$ and its
coherence coefficient is $\mu =1$.

\item The second example \cite{GHS1, GHS2, GHS3}, which is referred to as
the \textit{oscillator dictionary} $\mathfrak{D}_{O}$, is constructed using
the Weil representation of the finite symplectic group $SL_{2}\left( \mathbb{%
F}_{p}\right) $. The oscillator dictionary is of size approximately $p^{3}$
and its coherence coefficient is $\mu =4$.

\item The third example \cite{GHS1, GHS2, GHS3}, referred to as the \textit{%
extended oscillator dictionary} $\mathfrak{D}_{EO}$, is constructed using
the Heisenberg-Weil representation \cite{W, GH1}\ of the finite Jacobi
group, i.e., the semi-direct product $J\left( \mathbb{F}_{p}\right)
=SL_{2}\left( \mathbb{F}_{p}\right) \ltimes H\left( \mathbb{F}_{p}\right) $.
The extended oscillator dictionary is of size approximately $p^{5}$ and its
coherence coefficient is $\mu =4$.
\end{itemize}

The three examples of dictionaries we just described constitute reasonable
candidates for solving Problem \ref{deterministic_prb}: They are large in
the sense that $\left \vert \mathfrak{D}\right \vert \gg p,$ and empirical
evidences suggest (see \cite{AHSC} for the case of $\mathfrak{D}_{H})$ that
they might satisfy RIP with coefficients in the critical regime (\ref%
{critical_eq}). We summarize this as follows:

\begin{description}
\item[Question] \label{RIP_conj}Do the dictionaries $\mathfrak{D}_{H},%
\mathfrak{D}_{O}$ and $\mathfrak{D}_{EO}$ satisfy the RIP with coefficients $%
\delta _{1},\delta _{2}\ll 1$ and $n=\alpha \cdot p$, for some $0<\alpha <1$?
\end{description}

\section{Main results}

In this article we formulate a relaxed statistical version of RIP, called
statistical isometry property (SRIP for short) which holds for any
incoherent dictionary $\mathfrak{D}$ which is, in addition, a disjoint union
of orthonormal bases: 
\begin{equation}
\mathfrak{D=}\coprod_{x\in \mathfrak{X}}B_{x}\text{,}  \label{union_eq}
\end{equation}%
where $B_{x}=\left \{ b_{x}^{1},..,b_{x}^{p}\right \} $ is an orthonormal
basis of $\mathcal{H}$, for every $x\in \mathfrak{X}$.

\subsection{The statistical isometry property}

Let $\mathfrak{D}$ be an incoherent dictionary of the form (\ref{union_eq}).
Roughly, the statement is that for $S\subset \mathfrak{D}$, $\left \vert
S\right \vert =n$ with $n=p^{1-\varepsilon }$, for $0<\varepsilon <1$,
chosen uniformly at random, the operator norm $\left \Vert \mathbf{G}\left(
S\right) -Id_{S}\right \Vert $ is small with high probability. Precisely, we
have

\begin{theorem}[SRIP property \protect \cite{GH2}]
\label{SRIP_thm}For every $k\in 
%TCIMACRO{\U{2115} }%
%BeginExpansion
\mathbb{N}
%EndExpansion
$, there exists a constant $C\left( k\right) $ such that the probability 
\textbf{\ } 
\begin{equation}
P\left( \left \Vert \mathbf{G}\left( S\right) -Id_{S}\right \Vert \geq
p^{-\varepsilon /2}\right) \leq C\left( k\right) p^{1-\varepsilon k/2}.
\label{SRIP1_eq}
\end{equation}
\end{theorem}

The above theorem, \ in particular, implies that probability $P\left(
\left \Vert \mathbf{G}\left( S\right) -Id_{S}\right \Vert \geq p^{-\varepsilon
/2}\right) \rightarrow 0$ as $p\rightarrow \infty $ \ faster then $p^{-l}$
for any $l\in 
%TCIMACRO{\U{2115} }%
%BeginExpansion
\mathbb{N}
%EndExpansion
$.

\subsection{The statistics of the eigenvalues}

A natural thing to know is how the eigenvalues of the Gram operator $\mathbf{%
G}\left( S\right) $ fluctuate around $1$. In this regard, we study the
spectral statistics of the normalized error term 
\begin{equation*}
\mathbf{E}\left( S\right) \mathbf{=}\left( p/n\right) ^{1/2}\left( \mathbf{G}%
\left( S\right) -Id_{S}\right) .
\end{equation*}

Let $\rho _{\mathbf{E}\left( S\right) }=n^{-1}\sum_{i=1}^{n}\delta _{\lambda
_{i}}$ denote the spectral distribution of $\mathbf{E}\left( S\right) $
where $\lambda _{i}$, $i=1,..,n$, are the real eigenvalues of the Hermitian
operator $\mathbf{E}\left( S\right) $. The following theorem asserts that $%
\rho _{\mathbf{E}}$ converges in probability as $p\rightarrow \infty $ to
the Wigner semicircle distribution $\rho _{SC}\left( x\right) =\left( 2\pi
\right) ^{-1}\sqrt{4-x^{2}}\cdot \mathbf{1}_{\left[ 2,-2\right] }\left(
x\right) $ where $\mathbf{1}_{\left[ 2,-2\right] }$ is the characteristic
function of the interval $\left[ -2,2\right] $.

\begin{theorem}[Semicircle distribution \protect \cite{GH2}]
\label{Sato-Tate_thm}We have%
\begin{equation}
\lim_{p\rightarrow \infty }\rho _{\mathbf{E}}\overset{P}{=}\rho _{SC}\text{.}
\label{Sato-Tate_eq}
\end{equation}
\end{theorem}

\begin{remark}
A limit of the form (\ref{Sato-Tate_eq}) is familiar in random matrix theory
as the asymptotic of the spectral distribution of Wigner matrices.
Interestingly, the same asymptotic distribution appears in our situation,
albeit, the probability spaces are of a different nature (our probability
spaces are, in particular, much smaller).
\end{remark}

In particular, Theorems \ref{SRIP_thm}, \ref{Sato-Tate_thm} can be applied
to the three examples $\mathfrak{D}_{H}$, $\mathfrak{D}_{O}$ and $\mathfrak{D%
}_{EO}$, which are all of the appropriate form (\ref{union_eq}). Finally,
our result gives new information on a remark of
Applebaum-Howard-Searle-Calderbank \cite{AHSC} concerning RIP of the
Heisenberg dictionary.

\begin{remark}
For practical applications, it might be important to compute explicitly the
constants $C\left( k\right) $ which appears in (\ref{SRIP1_eq}). This
constant depends on the incoherence coefficient $\mu $, therefore, for a
fixed $p$, having $\mu $ as small as possible is preferable.
\end{remark}

\begin{acknowledgement}
It is a pleasure to thank our teacher J. Bernstein for his continuos
support. We are grateful to N. Sochen for many stimulating discussions. We
thank F. Bruckstein, R. Calderbank, M. Elad, Y. Eldar, R. Kimmel, and A.
Sahai for sharing with us some of their thoughts about signal processing. We
are grateful to R. Howe, A. Man, M. Revzen and Y. Zak for explaining us the
notion of mutually unbiased bases.
\end{acknowledgement}


\begin{thebibliography}{99}
\bibitem{AHSC} Applebaum L., Howard S., Searle S., and Calderbank R., Chirp
sensing codes: Deterministic compressed sensing measurements for fast
recovery.\textit{\ (Preprint, 2008). }

\bibitem{BDDW} Baraniuk R., Davenport M., DeVore R.A. and Wakin M.B., A
simple proof of the restricted isometry property for random matrices. 
\textit{Constructive Approximation, to appear (2007).}

\bibitem{BDE} Bruckstein A.M., Donoho D.L. and Elad M., "From Sparse
Solutions of Systems of Equations to Sparse Modeling of Signals and Images", 
\textit{to appear in SIAM Review }(2007).

\bibitem{C} Compressive Sensing Resources. Available at
http://www.dsp.ece.rice.edu/cs/.

\bibitem{Ca} Cand\`{e}s E. Compressive sampling. \textit{In Proc.
International Congress of Mathematicians, vol. 3, Madrid, Spain (2006).}

\bibitem{CRT} Cand\`{e}s E., Romberg J. and Tao T., Robust uncertainty
principles: exact signal reconstruction from highly incomplete frequency
information. \textit{Information Theory, IEEE Transactions on, vol. 52, no.
2, pp. 489--509 (2006).}

\bibitem{CT} Cand\`{e}s E., and Tao T., Decoding by linear programming. 
\textit{IEEE Trans. on Information Theory, 51(12), pp. 4203 - 4215 (2005). }

\bibitem{D} Donoho D., Compressed sensing.\textit{\ IEEE Transactions on
Information Theory, vol. 52, no. 4, pp. 1289--1306 (2006).}

\bibitem{DE} Donoho D.L. and Elad M., Optimally sparse representation in
general (non-orthogonal) dictionaries via l\_1 minimization. \textit{Proc.
Natl. Acad. Sci. USA 100, no. 5, 2197--2202} \textit{(2003).}

\bibitem{De} DeVore R. A., Deterministic constructions of compressed sensing
matrices. \textit{J. Complexity 23 (2007), no. 4-6, 918--925.}

\bibitem{DGM} Daubechies I., Grossmann A. and Meyer Y., Painless
non-orthogonal expansions. J\textit{. Math. Phys., 27 (5), pp. 1271-1283} 
\textit{(1986).}

\bibitem{EB} Elad M. and Bruckstein A.M., A Generalized Uncertainty
Principle and Sparse Representation in Pairs of Bases. \textit{IEEE Trans.
On Information Theory, Vol. 48, pp. 2558-2567} \textit{(2002).}

\bibitem{GH1} Gurevich S. and Hadani R., The geometric Weil representation . 
\textit{Selecta Mathematica, New Series, Vol. 13, No. 3. (December 2007),
pp. 465-481.}

\bibitem{GH2} Gurevich S. and Hadani R., Incoherent dictionaries and the
statistical restricted isometry property. \textit{Submitted to IEEE journal
of selected topics in signal processing - special issue on compressive
sensing (2008).}

\bibitem{GHS1} Gurevich S., Hadani R. and Sochen N., The finite harmonic
oscillator and its associated sequences. \textit{Proceedings of the National
Academy of Sciences of the United States of America, in press (2008).}

\bibitem{GHS2} Gurevich S., Hadani R., Sochen N., On some deterministic
dictionaries supporting sparsity . \textit{Special issue on sparsity, the
Journal of Fourier Analysis and Applications. To appear (2008). }

\bibitem{GHS3} Gurevich S., Hadani R., Sochen N., The finite harmonic
oscillator and its applications to sequences, communication and radar . 
\textit{IEEE Transactions on Information Theory, vol. 54, no. 9, September
2008. }

\bibitem{H} Howe R., Nice error bases, mutually unbiased bases, induced
representations, the Heisenberg group and finite geometries. \textit{Indag.
Math. (N.S.) 16 }, no. 3-4, 553--583 \textit{(2005).}

\bibitem{HCM} Howard S. D., Calderbank A. R. and Moran W. The finite
Heisenberg-Weyl groups in radar and communications. \textit{EURASIP J. Appl.
Signal Process.} \textit{(2006).}

\bibitem{HCS} Howard S.D., Calderbank A.R., and Searle S.J., A fast
reconstruction algorithm for deterministic compressive sensing using second
order Reed-Muller codes. \textit{CISS (2007).}

\bibitem{I} Indyk P., Explicit constructions for compressed sensing of
sparse signals. \textit{SODA (2008)}.

\bibitem{J} Jafarpour S., Efficient Compressed Sensing using Lossless
Expander Graphs with Fast Bilateral Quantum Recovery Algorithm. \textit{%
arXiv:0806.3799 (2008). }

\bibitem{JXHC} Jafarpour S., Xu W., Hassibi B., Calderbank R., Efficient and
Robust Compressive Sensing using High-Quality Expander Graphs. \textit{%
Submitted to the IEEE transaction on Information Theory (2008).}

\bibitem{XH} Xu W. and Hassibi B., Efficient Compressive Sensing with
Deterministic Guarantees using Expander Graphs. \textit{Proceedings of IEEE
Information Theory Workshop, Lake Tahoe (2007).}

\bibitem{S} Saligrama V., Deterministic Designs with Deterministic
Guarantees: Toeplitz Compressed Sensing Matrices, Sequence Designs and
System Identification. \textit{arXiv:0806.4958 (2008).}

\bibitem{Tr1} Tropp J.A., On the conditioning of random subdictionaries. 
\textit{Appl. Comput. Harmonic Anal., vol. 25, pp. 1--24, 2008.}

\bibitem{Tr2} Tropp J.A., Norms of random submatrices and sparse
approximation. \textit{Submitted to Comptes-Rendus de l'Acad\'{e}mie des
Sciences (2008).}

\bibitem{W} Weil A., Sur certains groupes d'operateurs unitaires. \textit{%
Acta Math. 111}, \textit{143-211 (1964).}
\end{thebibliography}
\end{document}